\documentclass[a4paper,10pt,twocolumn]{article}
\usepackage[footnotesize]{caption}
\usepackage[caption=false,font=footnotesize]{subfig}
\usepackage{mathptmx,graphicx,amsmath,amsfonts,amsthm,amssymb,url,cite}
\usepackage[margin={0.575in,0.8in}]{geometry}

\newcommand{\eps}{\varepsilon}

\newtheorem{theorem}{Theorem}

\newtheorem{lemma}{Lemma}
\newtheorem{corollary}{Corollary}

\begin{document}

\title{Discrete Distributions in the Tardos Scheme, Revisited}
\author{Thijs Laarhoven\footnote{T. Laarhoven and B. de Weger are with the Department of Mathematics and Computer Science, Eindhoven University of Technology, P.O. Box 513, 5600 MB Eindhoven, The Netherlands. \protect\\
E-mail: \{t.m.m.laarhoven,b.m.m.d.weger\}@tue.nl.} \and Benne de Weger\footnotemark[1]}
\date{\today}

\maketitle
\begin{abstract}
The Tardos scheme is a well-known traitor tracing scheme to protect copyrighted content against collusion attacks. The original scheme contained some suboptimal design choices, such as the score function and the distribution function used for generating the biases. \v{S}kori\'{c} et al.\ previously showed that a symbol-symmetric score function leads to shorter codes, while Nuida et al.\ obtained the optimal distribution functions for arbitrary coalition sizes. Later, Nuida et al.\ showed that combining these results leads to even shorter codes when the coalition size is small. We extend their analysis to the case of large coalitions and prove that these optimal distributions converge to the arcsine distribution, thus showing that the arcsine distribution is asymptotically optimal in the symmetric Tardos scheme. We also present a new, practical alternative to the discrete distributions of Nuida et al.\ and give a comparison of the estimated lengths of the fingerprinting codes for each of these distributions.
\end{abstract}


\section{Introduction}

To fight against copyright infringement, distributors of copyrighted content embed hidden watermarks in the data, creating a different version of the content for each user. Then, when a user distributes his copy and the distributor finds it, the distributor extracts the watermark from this copy and traces it to the guilty user. Assuming two versions can be created for every segment of the content, it is clear that with a binary search, $\ell \approx \log_2 n$ watermarked content segments suffice to find one pirate hidden among $n$ users.

Things become more complicated when several users \textit{collude}, and compare their differently watermarked copies to create a new version of the content that does not exactly match any of their copies. Assuming that for each segment of the data there are two different versions, and that in each segment the colluders output one of their received versions (known in the literature as the \textit{marking assumption}), it is impossible to trace $c \geq 2$ colluders deterministically (i.e., with no probability of error) with any fixed amount of segments. Fortunately, probabilistic schemes do exist that allow us to trace up to $c$ colluders with at most $\eps$ probability of error, for any given $c \geq 2$ and $\eps > 0$. One of the main objectives of research in this area is to construct such traitor tracing schemes, that allow us to trace colluders with as few segments $\ell$ as possible.


\subsection{Related work}

In 2003, Tardos~\cite{tardos03} showed that the optimal length of such codes (i.e., the number of segments needed) is of the order $\ell = d_{\ell} c^2 \ln(n/\eps_1)$ with $d_{\ell} = \Omega(1)$, where $\eps_1$ is an upper bound on the probability of catching one or more innocent users.\footnote{Note that $\eps_2$, commonly used for an upper bound on the probability of not catching any pirates, does not appear in the leading term of the code length for most practical values of $\eps_1$ and $\eps_2$.} In the same paper, Tardos gave a construction of a scheme with $d_{\ell} = 100$, which is widely known as the Tardos scheme. This shows that $d_{\ell} = \Theta(1)$ is optimal, and that the Tardos scheme has the optimal order code length. 

\sloppypar{Over the last ten years, improvements to the Tardos scheme have lead to a significant decrease in the code length parameter $d_{\ell}$. We previously showed~\cite{laarhoven12} that combining the symbol-symmetric score function of \v{S}kori\'{c} et al.~\cite{skoric08} with the improved analysis of Blayer and Tassa~\cite{blayer08} leads to an asymptotic code length constant of $d_{\ell} = \frac{1}{2}\pi^2 \approx 4.93$ for large $c$. For small coalitions, Nuida et al.~\cite{nuida09} showed that even smaller values $d_{\ell}$ can be obtained by combining the symmetric score function with the optimized, discrete distribution functions previously obtained by Nuida et al.~\cite{nuida07}. For large $c$, this lead to an asymptotic code length constant of about $d_{\ell} \approx 5.35$.}

Besides practical constructions of traitor tracing schemes, some papers have also studied absolute lower bounds on the asymptotic code lengths that any secure traitor tracing scheme must satisfy. Huang and Moulin~\cite{huang12} and Amiri and Tardos~\cite{amiri09} showed that for large $c$, the code length constant of any scheme must satisfy $d_{\ell} \geq 2$, but no practical constructions of schemes achieving this lower bound are known. Huang and Moulin did show that this lower bound is tight, and that in the related min-max game between the traitors and the tracer, the optimal pirate strategy is to use the interleaving attack, and the optimal tracing strategy is to use a Tardos-like code with biases distributed according to the arcsine distribution. Note that this does not say anything about specific schemes such as the Tardos scheme, for which the related min-max games are different and may lead to a completely different optimal pirate strategy and tracing strategy.


\subsection{Contributions and outline}

In this paper, we show that for large coalition sizes, the discrete distributions of Nuida et al.~\cite{nuida09, nuida07} converge to the arcsine distribution, thus proving that in the symmetric Tardos scheme, the arcsine distribution is asymptotically optimal. Together with results of \v{S}kori\'{c} et al.~\cite{skoric08} and us~\cite{laarhoven12}, this further implies that the asymptotic code length $\ell \sim \frac{1}{2} \pi^2 c^2 \ln(n/\eps_1)$ is optimal in the symmetric Tardos scheme. On the practical side, we present an alternative to the distributions of Nuida et al.\ with a simpler bias generation method, and conjecture that its performance is close to the performance of the distributions of Nuida et al. 

The outline of this paper is as follows. In Section~\ref{sec:pre} we describe the symmetric Tardos scheme, and different choices for the distribution function $F$ used in this scheme. In Section~\ref{sec:res} we state our results, and we devote Section~\ref{sec:1} to proving the main result. In Section~\ref{sec:approx} we present what we call discrete arcsine distributions, and in Section~\ref{sec:codelengths} we give a heuristic comparison of the lengths of the codes in the symmetric Tardos scheme when using these various distribution functions. Finally, in Section~\ref{sec:disc} we briefly discuss the results and mention a direction for future research.


\section{The symmetric Tardos scheme}
\label{sec:pre}

Before we describe the Tardos scheme, we introduce some more notation. The matrix $X = (X_{j,i})$, consisting of bits, is used to indicate which of the two versions of the $i$th content segment is assigned to user $j$, for each user $j \in \{1, \ldots, n\}$ and each segment $i \in \{1, \ldots, \ell\}$. We write $\vec{y} = (y_i)$ for the pirate output, consisting of $\ell$ bits.

The Tardos scheme roughly consists of two parts, which are outlined below. The scheme depends on appropriately chosen functions $F$ and $g$, and constants $\ell$ and $Z$. The first part of the scheme is performed before the content is distributed, and focuses on generating the code matrix $X$. The second part is performed once the pirates have output a forged copy $\vec{y}$ and this copy has been detected by the distributor, and focuses on finding the guilty users. 

\begin{itemize}
  \item[(1)] \textbf{Codeword generation} \\
  - For each $i$, generate $p_i \sim F$. \\
  - For each $i,j$, generate $X_{j,i} \sim \mathrm{Bernoulli}(p_i)$.
  \item[(2)] \textbf{Accusation algorithm} \\
  - For each $i,j$, compute $S_{j,i} = g(X_{j,i}, y_i, p_i)$. \\
  - For each $j$, accuse user $j$ if $\sum_{i=1}^{\ell} S_{j,i} > Z$.
\end{itemize}

This description is very general, and covers (almost) any known version of the Tardos scheme. The choice of $F$ and $g$, and the method to determine $\ell$ and $Z$, are what separates one scheme from another. In this paper we will focus on the class of \textit{symmetric} Tardos schemes, which means choosing $g$ as the symbol-symmetric score function of \v{S}kori\'{c} et al.~\cite{skoric08}:
\begin{align*}
g(X_{j,i}, y_i, p_i) = \begin{cases}
 +\sqrt{(1 - p_i)/p_i}, & \text{if $X_{j,i} = 1, y_i = 1$}, \\
 -\sqrt{(1 - p_i)/p_i}, & \text{if $X_{j,i} = 1, y_i = 0$}, \\
 -\sqrt{p_i/(1 - p_i)}, & \text{if $X_{j,i} = 0, y_i = 1$}, \\
 +\sqrt{p_i/(1 - p_i)}, & \text{if $X_{j,i} = 0, y_i = 0$}.
\end{cases}
\end{align*}
In this paper we will not go into detail about choosing $\ell$ and $Z$, but focus on the distribution function $F$. 


\subsection{Continuous arcsine distributions}

A common choice for the distribution function $F$ is the arcsine distribution with appropriate cutoffs. More precisely, we first compute a cutoff parameter $\delta_c > 0$, and we then use the distribution function $F_c$ defined on $[\delta_c, 1 - \delta_c]$ by:
\begin{align*}
F_c(p) = \frac{2 \arcsin \sqrt{p} - 2 \arcsin \sqrt{\delta_c}}{\pi - 4\arcsin \sqrt{\delta_c}}. \qquad (\delta_c \leq p \leq 1 - \delta_c)
\end{align*}
For $c = 10$, the distribution function $F_{10}$ is shown in Figure~\ref{fig:1}. For small values of $c$, the parameter $\delta_c$ has to be sufficiently large for a certain proof of security to work. For large $c$, the cutoff $\delta_c$ tends to $0$, and the distributions converge to the well-known arcsine distribution $F_{\infty}$, defined on $[0,1]$ by:
\begin{align*}
F_{\infty}(p) = \frac{2}{\pi} \arcsin \sqrt{p}. \qquad (0 \leq p \leq 1)
\end{align*}
With these continuous arcsine distribution functions, we previously showed~\cite{laarhoven12} that an asymptotic code length constant of $d_{\ell} \sim \frac{1}{2}\pi^2 \approx 4.93$ is optimal. For details, see \cite{laarhoven12}.


\subsection{Discrete Gauss-Legendre distributions}

Nuida et al.~\cite{nuida09, nuida07} showed that if the pirates aim to minimize their expected total score, the optimal distributions are in fact discrete distributions, and are related to Gauss-Legendre quadratures in numerical analysis. To define these distributions, we first need to introduce \textit{Legendre polynomials}. For $c \geq 1$, the $c$th Legendre polynomial is given by
\begin{align*}
P_c(x) = \frac{1}{2^c c!} \left(\frac{d}{dx}\right)^c (x^2 - 1)^c.
\end{align*}
This polynomial has $c$ simple roots on $(-1,1)$, which we will denote by $x_{1,c} < x_{2,c} < \ldots < x_{c,c}$. Now, the optimal distribution functions, for arbitrary $c$, are as follows. Here, optimal means that these distribution functions maximize the expected coalition score.

\begin{lemma} \cite[Theorem 3]{nuida07} \label{lem-nuida}
The optimal distribution to fight against $2c - 1$ or $2c$ colluders, is
\begin{align*}
F_{2c-1}(p) = F_{2c}(p) = \frac{1}{N_c} \sum_{k = 1}^c w_{k,c} H(p - p_{k,c}), \quad (0 \leq p \leq 1)
\end{align*}
where $N_c$ is a normalizing constant, $H$ is the Heaviside step function, and the points $p_{k,c}$ and weights $w_{k,c}$ are given by
\begin{align*}
p_{k,c} = \frac{x_{k,c} + 1}{2}, \quad w_{k,c} = \frac{2}{(1 - x_{k,c}^2)^{3/2} P_c'\left(x_{k,c}\right)^2}.
\end{align*}
\end{lemma}

\begin{figure}
\centering
\includegraphics[width=\columnwidth]{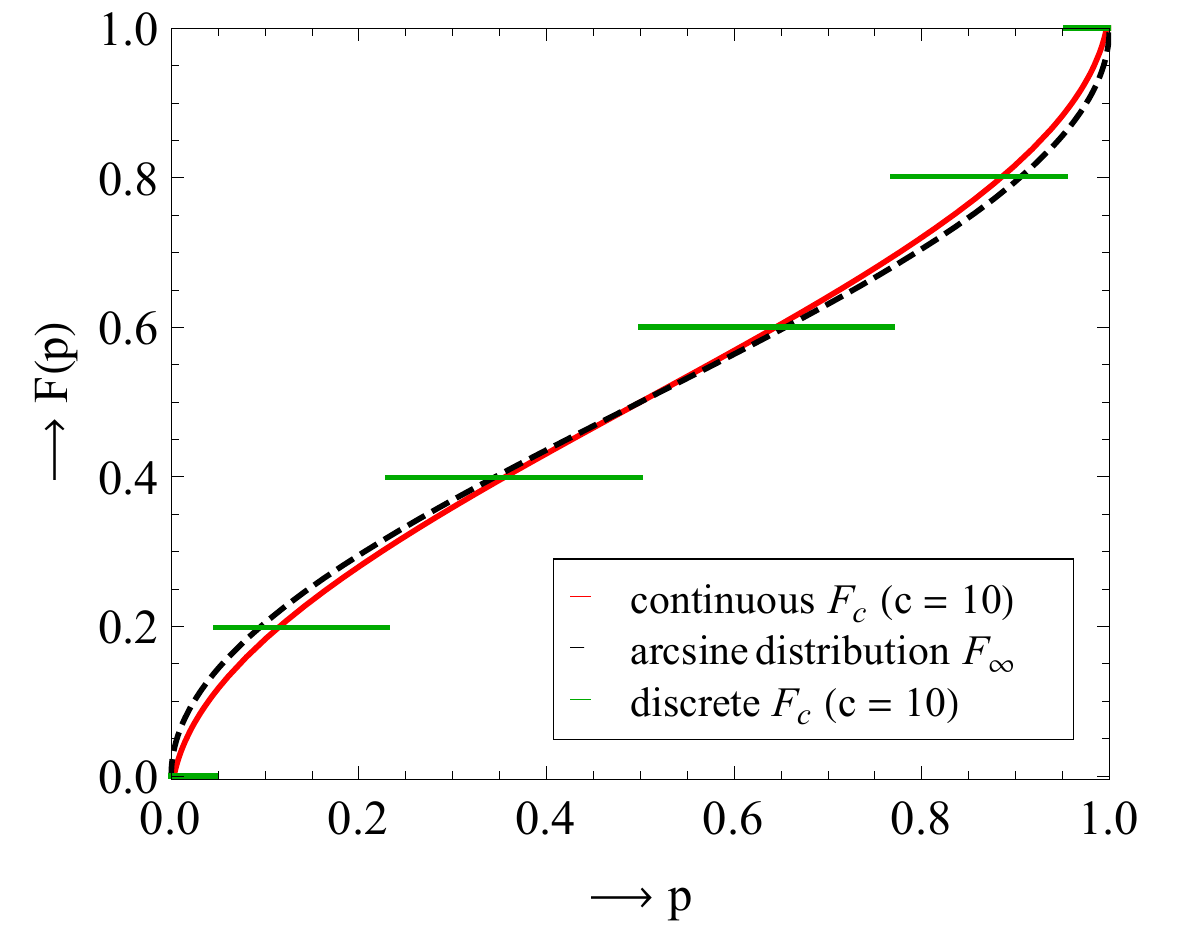}
\caption{The continuous arcsine distribution function with cutoff $\delta_{10} \approx 0.003$ (red) and the discrete Gauss-Legendre distribution function (green), both corresponding to the case $c = 10$. The dashed curve shows the arcsine distribution function. \label{fig:1}}
\end{figure}

The Gauss-Legendre distribution designed to resist $10$ colluders is shown in Figure~\ref{fig:1}. For small $c$, this construction gives much shorter codes than those obtained using the arcsine distributions with cutoffs. For large $c$, the code length parameter goes up, and Nuida et al.~\cite{nuida09} show that their results can be extended to a construction that asymptotically achieves a code length constant of the order $d_{\ell} \sim K \approx 5.35$. For details on this asymptotic result and Lemma~\ref{lem-nuida}, we refer the reader to~\cite{nuida09, nuida07}.

\subsection{The asymptotic optimum}

As we discussed above, the asymptotic code length constant obtained by Nuida et al.~\cite{nuida09} is slightly higher than the one obtained by us~\cite{laarhoven12}.  This means that the asymptotic result of Nuida et al.\ is not optimal. On the other hand, due to the Central Limit Theorem, the scores of innocent users (per segment) converge to the standard normal distribution with mean $0$ and variance $1$, while the total score of the coalition (per segment) converges to a normal distribution with some unknown mean $\tilde{\mu}$ and variance $\tilde{\sigma}^2 < \infty$, depending on the choice of $F$ and the pirate strategy. For fixed $\eps_2 > 0$ and large $c$, the only parameter that influences the asymptotic code length is the mean $\tilde{\mu}$, which is minimized by Nuida et al.'s choice of distribution functions. So we do expect that the asymptotic lengths of the codes in the symmetric Tardos scheme are minimized when using the distribution functions of Nuida et al. The fact that their asymptotic constant $d_{\ell} \approx 5.35$ is higher than the constant $d_{\ell} \approx 4.93$ which we showed to be sufficient for large $c$~\cite{laarhoven12}, suggests that Nuida et al.'s asymptotic analysis was not tight. Up until now, it was thus an open question what the best asymptotic distribution functions are in the symmetric Tardos scheme, as well as what their accompanying code length constants are. 


\section{Main results}
\label{sec:res}

We will prove that letting $c$ tend to infinity in the class of discrete distributions of Nuida et al.\ leads exactly to the arcsine distribution. This will be done by proving the following main result. 

\begin{theorem} \label{THM-1}
\sloppypar{Let the parameters $p_{k,c}$, $w_{k,c}$, and $N_c$ as in Lemma~\ref{lem-nuida}. Let $\alpha > 0$, and let $k$ satisfy $\alpha c < k < (1 - \alpha)c$. Then, as $c \to \infty$,}
\begin{align}
p_{k,c} &= \sin^2\left(\frac{\pi k}{2 c}\right) + o(1), \label{eq:p} \\
w_{k,c} &= \frac{\pi}{c} + o\left(\frac{1}{c}\right), \label{eq:w} \\
N_c &= \pi - o(1). \label{eq:N}
\end{align}
\end{theorem}

Note that except for the points near $0$ and $1$, corresponding to $k = o(c)$ or $k = c - o(c)$, the leading terms of the weights are all equal. But since these points in the `middle' carry $1 - o(1)$ weight (cf.~the proof of \eqref{eq:N}), the points near $0$ and $1$ have a negligible total weight. On the other hand, the points $p_{k,c}$ converge to the expected values of the corresponding order statistics of the arcsine distribution, i.e., the value $y$ corresponding to $F_{\infty}(y) = \frac{k}{c}$ is exactly $y = F_{\infty}^{-1}(\frac{k}{c}) = \sin^2(\frac{\pi k}{2c}) = p_{k,c} + o(1)$. Since asymptotically all these points have the same weight, after $k$ of the $c$ points we also have $F_{2c}(p_k^{(c)}) = \frac{k}{c} + o(\frac{1}{c})$ or $F_{2c}^{-1}(\frac{k}{c}) = p_{k,c} + o(1)$. Since the set of points $\{p_{k,c}\}_{k=1}^c$ is dense in $(0,1)$ when $c$ tends to infinity, these results imply that $F_{2c}(p) \to F_{\infty}(p)$ for each $p \in (0,1)$, proving that the arcsine distribution is asymptotically optimal in the symmetric Tardos scheme.

\begin{theorem}
In the symmetric Tardos scheme, the arcsine distribution is asymptotically optimal.
\end{theorem}

\v{S}kori\'{c} et al.~\cite[Section 6]{skoric08} previously showed that when using the arcsine distribution, due to the Central Limit Theorem the optimal code length inevitably converges to $\ell \to \frac{1}{2}\pi^2 c^2 \ln(n/\eps_1)$. So the following corollary is immediate.

\begin{corollary}
\sloppypar{In the symmetric Tardos scheme, the following code length is asymptotically optimal:}
\begin{align*}
\ell = \left(\frac{\pi^2}{2} + o(1)\right) c^2 \ln(n/\eps_1).
\end{align*}
\end{corollary}

In addition to these theoretical results, we present a new class of distribution functions, which can be obtained by discarding some of the order terms in Theorem~\ref{THM-1}. Compared to the Gauss-Legendre distributions, these distributions are much simpler, but still seem to achieve comparable code lengths. For details, see Sections~\ref{sec:approx} and \ref{sec:codelengths}.


\section{Proof of Theorem~\ref{THM-1}}
\label{sec:1}

\eqref{eq:p}: Let $\theta_{k,c} = \arccos(x_{k,c})$. From \cite[Eq.~(22.16.6)]{abramowitz72} 
we have
\begin{align}
\theta_{k,c} = \left(\frac{4(c - k) + 3}{4c + 2}\right)\pi + o(1) = \pi - \frac{\pi k}{c} + o(1). \label{eq:theta}
\end{align}
Using $\cos(\pi - \phi) = 2 \sin^2 (\frac{\phi}{2}) - 1$ for $\phi \in \mathbb{R}$, we get
\begin{align*}
x_{k,c} = \cos\left(\pi - \frac{\pi k}{c} + o(1)\right) 
= 2 \sin^2\left(\frac{\pi k}{2c}\right) - 1 + o(1).
\end{align*}
Since $p_{k,c} = \frac{1}{2}(x_{k,c} + 1)$, Equation~\eqref{eq:p} follows. \\
\eqref{eq:w}: Combining \cite[Eq.~(15.3.1)]{szego75} and \cite[Eq.~(15.3.10)]{szego75}, and using $2 \sin(\frac{\theta_{k,c}}{2}) \cos(\frac{\theta{k,c}}{2}) = \sin(\theta_{k,c})$, we get
\begin{align*}
\frac{2}{(1 - x_{k,c}^2) P_c'(x_{k,c})^2} = \frac{\pi}{c} \sin(\theta_{k,c}) + o\left(\frac{1}{c}\right).
\end{align*}
Dividing both sides by $\sqrt{1 - x_{k,c}^2} = \sin \theta_{k,c}$ leads to \eqref{eq:w}. \\
\eqref{eq:N}: The Gauss-Legendre quadrature rule~\cite[Eq.~(25.4.29)]{abramowitz72} states that for analytic functions $f$, there exist constants $A_c > 0$ and $\xi \in (-1,1)$, with 
\begin{align*}
\int_{-1}^1 f(x) dx = \sum_{k=1}^c \frac{2 f(x_{k,c})}{(1-x_{k,c}^2) P_c'(x_{k,c})^2} + A_c f^{(2c)}(\xi).
\end{align*}
Let $f(x) = (1 - x^2)^{-1/2}$. Then we have $f^{(2c)}(x) > 0$ for all $x \in (-1,1)$, so in particular $f^{(2c)}(\xi) > 0$. So it follows that
\begin{align}
\pi = \int_{-1}^1 \frac{dx}{\sqrt{1 - x^2}} > \sum_{k = 1}^c w_{k,c} = N_c. \label{eq:cg}
\end{align}
On the other hand, from \eqref{eq:w} and $w_{k,c} > 0$ for all $k$, we have
\begin{align*}
N_c > \sum_{k=o(c)}^{c - o(c)} w_{k,c} = (c - o(c))\left(\frac{\pi}{c} + o\left(\frac{1}{c}\right)\right) = \pi - o(1).
\end{align*}
So $\pi - o(1) < N_c < \pi$, which proves \eqref{eq:N}.


\section{Discrete arcsine distributions}
\label{sec:approx}

By making a slight refinement to \eqref{eq:p} using \eqref{eq:theta}, we get that for large $c$ and almost all values of $k$, the parameters of the optimal distributions satisfy
\begin{align*}
p_{k,c} \approx \sin^2\left(\frac{4k - 1}{8c + 4}\pi\right), \quad w_{k,c} \approx \frac{\pi}{c}, \quad N_c \approx \pi.
\end{align*}
To get the exact values of these parameters for large $c$ requires quite some effort, so in practice one may consider using an approximation of these distributions. An obvious approximation to the above weights and points would be
\begin{align*}
p'_{k,c} = \sin^2\left(\frac{4k - 1}{8c + 4}\pi\right), \quad w'_{k,c} = \frac{\pi}{c}, \quad N'_c = \pi.
\end{align*}
Generating biases $p$ from the associated distribution function is equivalent to drawing $r$ uniformly at random from $\{\frac{3\pi}{8c + 4}, \frac{7\pi}{8c + 4}, \ldots, \frac{\pi}{2} - \frac{3\pi}{8c + 4}\}$, and setting $p = \sin^2(r)$. Note that if we were to draw $r$ uniformly at random from the complete interval $[0, \frac{\pi}{2}]$, this would correspond to the arcsine distribution, while drawing $r$ uniformly at random from $[\arcsin(\sqrt{\delta}), \frac{\pi}{2} - \arcsin(\sqrt{\delta})]$ corresponds to the arcsine distribution with cutoff $\delta$. So these distributions may be appropriately called \textit{discrete arcsine distributions}, and needless to say, for large $c$ these distributions also converge to the arcsine distribution.

\textit{Remark.} Interestingly, slightly different parameters,
\begin{align*}
p''_{k,c} = \sin^2\left(\frac{4k - 2}{8c}\pi\right), \quad w''_{k,c} = \frac{\pi}{c}, \quad N''_c = \pi,
\end{align*}
\hyphenation{Che-by-shev}
correspond exactly to the parameters of the so-called \textit{Cheby-shev-Gauss quadratures} \cite[Eq.~(25.4.38)]{abramowitz72}. These quadratures allow one to approximate integrals of the form
\begin{align}
\int_{-1}^1 \frac{g(x)}{\sqrt{1 - x^2}} dx \approx \sum_{k=1}^c w''_{k,c} g(x''_{k,c}), \label{eq:cg2}
\end{align}
where $x''_{k,c} = 2p''_{k,c} - 1$.\footnote{Also note the resemblance between Equations~\eqref{eq:cg2} and \eqref{eq:cg}, when $g(x) \equiv 1$ is the constant function with value $1$.} The distribution functions generated by these weights and points are very similar to the discrete arcsine distributions described above. The main difference seems to be the ``cutoff'', which would be about a third smaller (i.e., $\frac{2\pi}{8c}$ compared to $\frac{3\pi}{8c + 4}$). Since these distributions are worse approximations of the optimal Gauss-Legendre distributions, it seems that the discrete arcsine distributions are a better alternative.


\section{Estimating Codelengths}
\label{sec:codelengths}

Let us now try to give a qualitative comparison of the several classes of discrete and continuous distribution functions, in terms of code lengths. Since the (tails of) distributions of user scores are hard to estimate, and known proof methods are not tight, we will only give a heuristic estimate of the code lengths. Getting more accurate estimates remains an open problem.

Assuming that the scores of users are Gaussian, we can get a reasonable estimate for the optimal code length parameter as $d_{\ell} \approx 2/\tilde{\mu}^{2}$, where $\tilde{\mu}$ is the expected average pirate score per content segment \cite[Corollary 2]{skoric08}. In the case of the discrete distributions of Nuida et al., $\tilde{\mu}$ does not depend on the pirate strategy, so we can compute $\tilde{\mu}$ exactly. For the arcsine distributions with cutoffs and the discrete arcsine distributions, $\tilde{\mu}$ does depend on the pirate strategy, but by considering the attack that minimizes $\tilde{\mu}$ we can obtain lower bounds on $\tilde{\mu}$. 

\begin{figure}
\centering
\includegraphics[width=\columnwidth]{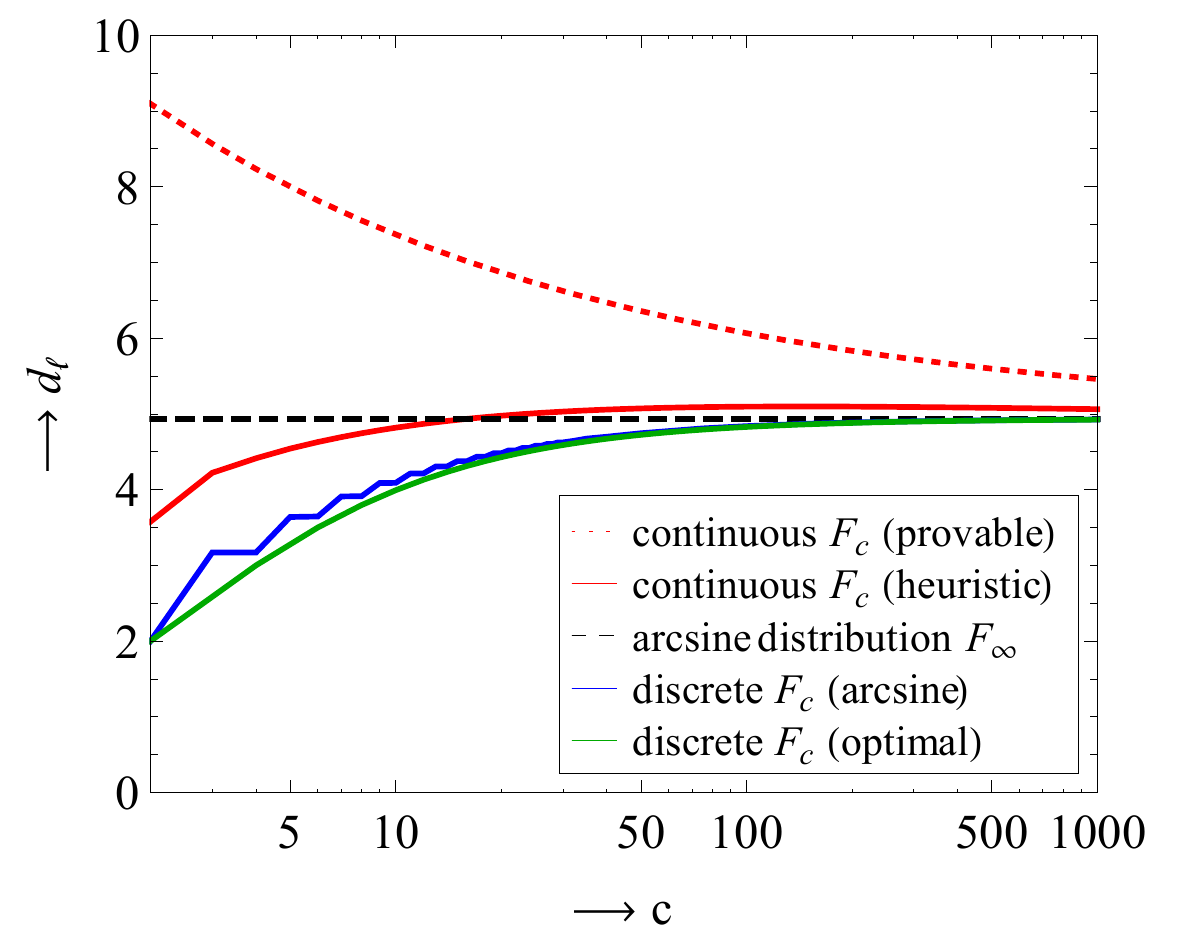}
\caption{Estimates of the code length parameters $d_{\ell}$ for several types of distribution functions $F$. The dashed line shows the asymptotic optimal value $d_{\ell} = \frac{1}{2}\pi^2$, corresponding to the arcsine distribution $F_{\infty}$. \label{fig:2}}
\end{figure}

Figure~\ref{fig:2} shows the resulting estimates of $d_{\ell}$, as well as the provable upper bounds on $d_{\ell}$ of \cite{laarhoven12} (for constant $\eps_2$). Note that the heuristic estimates for the continuous distributions are based on the arcsine distributions with cutoffs optimized for the proof technique of \cite{laarhoven12}. A different optimization of the cutoffs would lead to different (smaller) values of $d_{\ell}$.

\section{Conclusion}
\label{sec:disc}

We have shown that the optimal discrete distributions of Nuida et al.\ converge to the arcsine distribution, hence showing that the arcsine distribution is asymptotically optimal in the symmetric Tardos scheme. This connects the world of the discrete distributions to the world of the continuous distributions, as both converge to the same distribution. 

In practice, the question remains which distribution function to choose. In the static Tardos scheme, choosing one of the discrete distributions seems logical, as this may drastically reduce the length of the fingerprinting code. On the other hand, when $c$ is unknown, using the continuous distribution functions with cutoffs may also have its benefits, since if the coalition is slightly larger than expected, the scheme still has a good chance of catching the pirates. Recently, it was shown that the Tardos scheme can also be extended to the \textit{dynamic} traitor tracing setting, allowing efficient tracing of pirates when the colluders broadcast their forged copy in real-time \cite{laarhoven11c, laarhoven12c}. The construction of the universal Tardos scheme in \cite{laarhoven11c} uses the fact that the continuous distributions are very similar for different values of $c$, so in this setting it seems that the continuous arcsine distributions are also more practical.

An interesting open problem is what happens when the number of versions per content segment increases from $2$ to $q$. Recent results show \cite{huang12b} that with unlimited computing power, the optimal asymptotic code length decreases linearly in $q$. This suggests that the optimal length of $q$-ary Tardos codes possibly decreases linearly in $q$ as well. \v{S}kori\'{c} et al.~\cite{skoric08} analyzed a natural generalization of the Tardos scheme to the $q$-ary setting, but did not obtain this linear decrease in $q$ in their code lengths. The question remains whether their construction is suboptimal (and if so, whether this has to do with the choice of $F$ or the choice of $g$), or if approaching the fingerprinting capacity for higher $q$ with a single decoder traitor tracing scheme is simply impossible.

\section{Acknowledgments} 
We thank Jeroen Doumen, Wil Kortsmit, Jan-Jaap Oosterwijk, Georg Prokert, Berry Schoenmakers, and Boris \v{S}kori\'{c} for valuable discussions. We would also like to thank the anonymous reviewers for their useful comments.

\end{document}